# The $\mathbb{R}$-Algebra of Quasiknowledge and Convex Optimization

*2022-12-08*
*Duyal Yolcu*
*https://github.com/qudent*

This article develops a convex description of a classical or quantum learner's or agent's state of knowledge about its environment, presented as a convex subset of a commutative R-algebra. With caveats, this leads to a generalization of certain semidefinite programs in quantum information (such as those describing the universal query algorithm dual to the quantum adversary bound, related to optimal learning or control of the environment) to the classical and faulty-quantum setting, which would not be possible with a naive description via joint probability distributions over environment and internal memory. More philosophically, it also makes an interpretation of the set of reduced density matrices as "states of knowledge" of an observer of its environment, related to these techniques, more explicit. As another example, I describe and solve a "formal differential equation of states of knowledge" in that algebra, where an agent obtains experimental data in a Poissonian process, and its state of knowledge evolves as an exponential power series.

However, this framework currently lacks impressive applications, and I post it in part to solicit feedback and collaboration on those. In particular, it may be possible to develop it into a new framework for the design of experiments, e.g. the problem of finding maximally informative questions to ask human labelers or the environment in machine-learning problems. The parts of the article not related to quantum information don't assume knowledge of it.

# Contents

# 1 Outline

The article starts with a more detailed description of the motivation, and the problems arising when describing states of knowledge as joint probability distributions, and how the approach here is different (Section 2). In Section 3 - Section 7, I develop the concrete situations of classical, pure quantum, and mixed quantum physics and obtain structures $\mathcal{S}^\pm_{\text{class}}$, $\mathcal{S}^\pm_{\text{quant}}$, $\mathcal{S}^\pm_{\text{decoh}}$. More precisely, Section 3 defines the algebras of knowledge for classical and pure quantum theory in terms of generators and relations, Section 4 defines sets, Section 5 - Section 6 structure on that sets, and Section 7 discusses the equivalence relation and basic algebraic structure of $\mathcal{S}^\pm$.

Subsection 8.1 is a basic example, Subsection 8.2 introduces a new concept: Formal differential equations of knowledge.

Section 9 - Section 11 discuss convex optimization problems to help understand the "statics" and "dynamics" of states of knowledge better — answering questions like

- "What functions of the environment can I calculate with a given maximal error probability and state of knowledge?",
- "How similar are two states of knowledge?",
- "Which states of knowledge are achievable by an agents in $N$ interaction rounds with its environment?", or
- "How many steps does an agent need to accomplish a task?"

Finally, Section 13 gives a literature overview.

Note that much of this text is only relevant for the case of quantum computation involving quantum channels. A reader who wants to focus on the classical case may read Subsection 3.1, and then go to Section 8 and the following ones directly; the pure-quantum case is discussed in Subsection 3.2.

## 2 Intuition and motivation

Major developments in quantum computational complexity theory are the *adversary method* [Ambainis, Belovs], and more generally, semidefinite programming/convex optimization methods in quantum information [SikoraVarvitsiotis]. The aim of this note is to generalize these methods to classical information theory, as well as quantum information theory involving mixed states. This leads to a general notion of an agent's knowledge about its environment, and the actions it can perform to learn about and manipulate that environment, which I'll call "state of knowledge". For example, this framework could be applicable if one trains a machine learning model and wants to determine the "most useful questions" to ask human labelers or the real-world environment, if compute is much cheaper than gathering that data. I'll focus on situations where the agent has unbounded internal computational power, and is only restricted by the interface with the environment. Turning this framework into a realistic machine learning system — for example, involving neural networks learning representations of the algebraic structures involved — is a natural topic of future work.

One can compare this approach to generalized probabilistic theories, known in the foundations of quantum physics — a family of mathematical structures that generalize both quantum physics and classical probability theory. To a limited degree, they allow reasoning about quantum and classical systems using the same algebraic manipulations as well. However, the approach here does not describe a generalized probabilistic theory.

In probability theories, one would describe the "state of knowledge" by a joint probability distribution over the environment's and the agent's internal memory states. The approach I present is different in **two basic ways:**

1. **Equivalent joint probability distributions are the same state of knowledge:** In terms of "how much the agent knows about its environment", many joint probability distributions should be considered equivalent — for example, if the environment and the internal memory are both described by one bit, the joint probability matrices[1] $\begin{pmatrix} 0.5 & 0 \\ 0 & 0.5 \end{pmatrix}$ and $\begin{pmatrix} 0 & 0.5 \\ 0.5 & 0 \end{pmatrix}$ both correspond to the agent having complete knowledge of the environment. We will formalize this idea by an equivalence relation over which we take equivalence classes — essentially letting states be equivalent if the agent can transform them into each other without interacting with the environment. Then only distinct equivalence classes correspond to distinct "states of knowledge".
2. **The addition used for convex mixtures corresponds to a direct, not entrywise sum on the joint probability matrices:** A basic feature of classical probability distributions and mixed quantum states is *convexity*: If $A$ and $B$ are valid probability distributions, transition matrices or similar, and $0 \leq p \leq 1$, an object denoted by $c_p(A, B) := pA + (1-p)B$ exists and is to be interpreted as "the result of choosing $A$ with probability $p$ and $B$ with probability $1 - p$". In probability theory, $A$ and $B$ are some vectors or matrices, $pA$ denotes the **entrywise** product with $p$, and the $+$ sign denotes the **entrywise sum** of these objects.

In our notion of "states of knowledge", a "convex mixture" $c_p(A, B) = pA + (1-p)B$ should correspond to the agent "being in state of knowledge $A$ with probability $p$, and in state of knowledge $B$ with probability $1 - p$." **That wouldn't work if we just took elementwise convex combinations of the joint probability distributions:** Reusing the example from 1.,
$$\frac{1}{2}\begin{pmatrix} 0.5 & 0 \\ 0 & 0.5 \end{pmatrix} + \frac{1}{2}\begin{pmatrix} 0 & 0.5 \\ 0.5 & 0 \end{pmatrix} = \begin{pmatrix} 0.25 & 0.25 \\ 0.25 & 0.25 \end{pmatrix}$$
corresponds to a state of complete ignorance even though the constituent matrices correspond to perfect knowledge. We need the system to remember *which* constituent matrix was chosen. To do this, we use **direct sums:** Interpreting $\oplus$ as a direct sum of the columns,
$$\frac{1}{2}\begin{pmatrix} 0.5 & 0 \\ 0 & 0.5 \end{pmatrix} \oplus \frac{1}{2}\begin{pmatrix} 0 & 0.5 \\ 0.5 & 0 \end{pmatrix} = \begin{pmatrix} 0.25 & 0 & 0 & 0.25 \\ 0 & 0.25 & 0.25 & 0 \end{pmatrix}.$$

We interpret the right-hand side's memory space as follows: It contains an additional bit, which is set to 0 if the knowledge's encoding corresponds to that of the left summand, and 1 if it corresponds to that of the right summand. It will turn out that the right-hand side matrix is the state of complete knowledge as well — in other words, another representative of the equivalence class that contains $\begin{pmatrix} 0.5 & 0 \\ 0 & 0.5 \end{pmatrix}$ and $\begin{pmatrix} 0 & 0.5 \\ 0.5 & 0 \end{pmatrix}$. This is as it should be, as having complete knowledge in either of two ways, with probability $1/2$ each, is equivalent to having complete knowledge with probability 1.

So in our formal algebraic structure of "states of knowledge" describing classical physics, multiplication with a nonnegative scalar is defined elementwise, and addition is implemented by a direct sum. If we find a way to subtract states of knowledge as well (or multiply with a negative scalar), we have all the ingredients needed for a real vector space (if everything fulfills the axioms). To make states subtractable, we use a *Grothendieck group construction*, which is a consistent way to add enough negatives to an algebraic structure that everything has an inverse (just like negative numbers were abstracted from natural numbers as "numbers that don't actually exist, but are useful for calculations"). In the end, the physical states of knowledge are a convex subset of a vector space.

## 2.1   Contributions and applications

The algebraic structure defined allows us to do the following:

1. As mentioned in the beginning, we will be able to describe the problem of finding an optimal strategy to manipulate or learn about the environment as a **convex optimization problem** — and hopefully use duality to find lower bounds for the agent's ability to do so. We can describe results from quantum complexity theory, in particular Barnum-Saks-Szegedy [Barnum-Saks-Szegedy] and the related quantum adversary bound-universal query algorithm duality [Ambainis], [Belovs-Yolcu], [Yolcu], in our formalism — and plugging in the definitions, the convex optimization problems therein generalize to classical or decoherent-quantum situations. **Caveats** are that
    - the dimension of the involved vector spaces in the classical or decoherent-quantum case is generally infinite, in contrast to the coherent-quantum case,
    - I currently don't see an analogue of Slater's strong duality condition, or in fact simple descriptions of the dual problems.

2. Leaving mathematical rigour a bit (as we don't define infinite sums or limits in this note), we can also describe **formal differential equations** and construct power series that describe e.g. the evolution of a learner's knowledge over time when it observes a Poisson process generating experimental data. This will involve a notion of multiplication of "states of knowledge", corresponding to an agent having access to multiple states of knowledge in parallel (e.g. multiple observations gathered at different times). This notion gives the states of knowledge the structure of an $\mathbb{R}$-algebra (which can be defined as a vector space with additional structure).

# 3 Definition by generators and relations

In this section, I briefly present an alternative definition of classical or pure-quantum (not mixed-quantum) states of knowledge in terms of formal linear combinations of probability vectors or wavefunctions, subject to relations.

So consider a finite set $E$, representing possible states of the environment.

## 3.1 Classical states of knowledge

We define the $\mathbb{R}$-vector space $\mathcal{S}_{\text{class}}^{\pm E}$ as the vector space of **formal** (finite) linear combinations of vectors of nonnegative real numbers over $E$, $\vec{p} \in \mathbb{R}_{\geq 0}^{E}$ : An element $S \in \mathcal{S}_{\text{class}}^{\pm E}$ can be written as

$$S = \lambda_1 \left[\vec{p_1}\right] + \lambda_2 \left[\vec{p_2}\right] + \ldots + \lambda_N \left[\vec{p_N}\right],$$

where we write $\left[\vec{p_i}\right]$ when considering the vectors as elements of $\mathcal{S}_{\text{class}}^{\pm E}$. For any $q \geq 0$ and $\vec{p} \in \mathbb{R}_{\geq 0}^{E}$, we introduce a relation, i.e. linear dependency, of

$$q\left[\vec{p}\right] = \left[q\vec{p}\right],$$

and no other dependencies - in particular, in general,

$$\left[\vec{p_1} + \vec{p_2}\right] \neq \left[\vec{p_1}\right] + \left[\vec{p_2}\right].$$

The set $\mathcal{S}^E \subset \mathcal{S}^{\pm E}$ consists of those elements that can be written without negative coefficients.

> ⓘ **The vector space corresponds to an agent's joint probability matrix, modulo transformations**
>
> Such a positive linear combination should be interpreted as a joint probability matrix over environmental and internal memory states, $\left(\lambda_1 \vec{p_1}, \lambda_2 \vec{p_2}, \ldots\right)$ — and the freedom in choosing representatives and rearranging the generators corresponds exactly to the agent's freedom in reversibly acting on its own internal memory states:
>
> - Rearranging the generators corresponds to relabeling the internal memory states,
> - The linear dependency $q\left[\vec{p}\right] = \left[q\vec{p}\right]$ doesn't change the matrices,
> - "splitting" generators as in $\lambda_1 \left[\vec{p}\right] + \lambda_2 \left[\vec{p}\right] = (\lambda_1 + \lambda_2) \left[\vec{p}\right]$, or vice versa, corresponds to a probabilistic process in which the agent generates an additional random bit in case it is in the appropriate memory state, which is reversible by deleting that bit.
>
> In conclusion, each normalized probability vector from the generating set corresponds to a prior probability distribution of environmental states — and a convex mixture should be interpreted as a probability distribution over the agent having a certain posterior probability distribution over $E$.

Then

2. We define the trace, $\text{tr}\colon \mathcal{S}^{\pm E} \to \mathbb{R}$, as the linear map that maps each generating vector as $[\vec{p}] \to \sum_{e \in E} p_e$,
3. For a bipartite environment $E' = E \times X$, the partial trace $\text{tr}_X \colon \mathcal{S}^{\pm E \times X} \to \mathcal{S}^{\pm E}$ maps a multipartite probability vector to a sum of its individual columns: If we write $\vec{p} \in \mathbb{R}_{\geq 0}^{E \times X}$ as a vector of vectors $\left(\vec{p_x}\right)_{x \in X}$ with each $\vec{p_x} \in \mathbb{R}_{\geq 0}^{E}$, $\text{tr}_X \vec{p} := \sum_{x \in X} \left[\vec{p_x}\right]$. We again extend this so that $\text{tr}_X$ is linear, and corresponds to absorbing part of the environment into the agent's internal memory.

4. We generate a preorder that captures the agent's ability to "forget" some things, i.e. start to treat different states as one: For any $\vec{p_1}, \vec{p_2}$, we require

$$0 \leq \left[\vec{p_1} + \vec{p_2}\right] \leq \left[\vec{p_1}\right] + \left[\vec{p_2}\right]$$

as well as that $\leq$ is a preorder compatible with the vector space structure as defined in [Wikipedia](#): For all $x \leq y \in \mathbb{S}^{\pm E}$, we require $x + z \leq y + z$ for all $z \in \mathbb{S}^{\pm E}$ and $\lambda x \leq \lambda y$ for all $\lambda \geq 0 \in \mathbb{R}$.

So we define $S_1 \leq S_2$ if and only if $S_1 \leq S_2$ can be deduced in a finite number of steps from the requirement above, the preorder axioms, and the vector space compatibility. We can prove that this is actually a partial order, i.e. $S_1 \leq S_2 \leq S_1$ implies $S_1 = S_2$ :

1. First, if $S_1 \leq S_2 \leq S_1$, then the transformations implying these inequalities can't include a nontrivial instance of $0 \leq [\vec{p}]$, as this would strictly decrease the trace and imply $\text{tr} S_1 < \text{tr} S_1$.
2. Define the "expected entropy" $S(S)$ as a linear map that maps normalized probability vectors to their Shannon

entropies (and formal linear combinations of these to the appropriate linear combination),
3. By strict concavity of entropy, this quantity will strictly increase through each "forgetting transformation" that leads to a different state,
4. Therefore, $S_1 \leq S_2 \leq S_1$ and $S_1 \neq S_2$ would imply that $S(S_1) < S(S_2) < S(S_1)$, a contradiction.
5. We turn this into an associative, commutative ordered algebra by defining multiplication elementwise:

$$\left[\overrightarrow{p_1}\right]\left[\overrightarrow{p_2}\right] := \left[\overrightarrow{p_3}\right],$$

$$(p_3)_e := (p_1)_e (p_2)_e,$$

and again extending bilinearly to the rest of the vector spaces. This corresponds to the agent having several sources of information at once, see Subsection 8.1 for an example.

## 3.2 Pure quantum theory

is treated similarly: The $\mathbb{R}$-vector space $\mathcal{S}_{\text{quant}}^{\pm E}$ is generated by nonnormalized wavevectors over $E$ (i.e. elements $\Psi \in \mathbb{C}^E$), the rescaling relation that $[\alpha \Psi] = |\alpha|^2 [\Psi]$ for any $\alpha \in \mathbb{C}$ and $\Psi \in \mathbb{C}^E$, and furthermore a **polarization identity:** For any $\Psi_1, \Psi_2 \in \mathbb{C}^E$,

$$[\Psi_1] + [\Psi_2] = \frac{1}{2}[\Psi_1 + \Psi_2] + \frac{1}{2}[\Psi_1 - \Psi_2].$$

In analogy to the classical situation, one can interpret a sum (and by the rescaling identity, any positive linear combination) of $[\Psi_i]$ as an agent with several internal memory states, each corresponding to a certain amplitude distribution over possible environmental states. The linear combination corresponds to a bipartite wavevector over environment and internal memory (with each internal memory state a summand). We will now see that the equivalence relations correspond precisely to equivalence of those wavevectors that the quantum computer can transform into each other by unitary transformations acting on the internal memory (considered as a subset of a countably infinite-dimensional internal memory space):

> 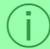 **The vector space corresponds to an agent's joint amplitude matrix, modulo transformations**
>
> Vector space representatives that differ by a sum of $[0]$, or are constructed by adding generators in a different order, are equivalent. Correspondingly on the side of bipartite wavevectors, wavevectors are equivalent that differ by relabeling memory states (equivalent to a permutation unitary); we also consider wavevectors as equivalent that are related by adding and/or removing memory states with only 0 amplitudes.
>
> Now, consider a representative of the vector space as a sum of pairwise different generators not containing $[0]$, with prefactor 1. To show the quantum computer can transform them by operations local to the memory space, we need to reproduce the effect of the polarization identity, rescaling, relabeling memory states, and splitting/joining states in the form of $\left[\sqrt{p}\Psi\right] + \left[\sqrt{1-p}\Psi\right] = [\Psi] = [\Psi] + [0]$. On the side of the bipartite states, these transformations corresponds to a sequence of Hadamard (on the corresponding two levels), diagonal phase, permutation unitaries, and appropriate rotations.
>
> Conversely, any isometry acting on a bipartite wavevector's memory can be decomposed into such operations: We add a sufficient number of all-0 memory states and extend the isometry to a unitary. Then, by a standard result (see e.g. [NielsenChuang], Section 4.5.1: "Two-level unitary gates are universal"), we can decompose the unitary to a sequence of two-level unitaries acting on combinations of memory states. In turn, we can decompose these into sequences of

> Hadamards, rotations around $Z$, and global phase shifts - all of which correspond to equivalence relations between vector space representatives.

We skip defining partial trace, partial order etc. explicitly and refer to the general discussion starting in Section 4; the results are similar to the previous ones for classical states.

### 3.2.1 Pure-quantum states of knowledge are reduced density matrices

In quantum probability theory, the **states of knowledge correspond to the reduced density matrices** of these bipartite states on the subspace corresponding to the environment. This is because for any bipartite quantum system $\mathcal{E} \otimes \mathcal{M}$, the set of nonnormalized density matrices on $\mathcal{E}$ is in 1-to-1 correspondence with the set of equivalence classes of pure states on $\mathcal{E} \otimes \mathcal{M}$, modulo local unitary transformations on $\mathcal{M}$: Every density matrix can be purified, local unitaries on the purifying space don't change the reduced density matrix, and all purifications of a given reduced density matrix are related by a local unitary [NielsenChuang].

One can verify that addition and multiplication of states of knowledge corresponds to entrywise addition and multiplication of these density matrices, our partial trace (defined analogously to the classical probabilistic case) is the partial trace of density matrices, and $S_1 \leq S_2$ iff $0 \leq S_2 - S_1$, i.e. iff $S_2 - S_1$ corresponds to a positive semidefinite matrix itself.

### 3.2.2 Relation to quantum query algorithms and the adversary bound

A discussion of a quantum or classical algorithm typically involves tracking the quantum computer's wavefunction for the various inputs.

We can formulate this in the framework of bipartite wavefunctions and states of knowledge: If we let the environment $E$ be the set of possible inputs, a bipartite, nonnormalized wavefunction $\sum_{e \in E} |e\rangle \otimes |\Psi_e\rangle \in \mathcal{E} \otimes \mathcal{M}$ may encode all of these states at once. By the preceding discussion, the corresponding state of knowledge captures precisely that state modulo permissible transformations by a quantum query algorithm as described e.g. in [Belovs-Yolcu].

The transpose of the wavefunction's reduced density matrix on $\mathcal{E}$ (which we know corresponds to the state of knowledge) equals the *Gram matrix*, i.e. the matrix of complex inner products between the individual states. Tracking feasible Gram matrices is the idea behind the so-called *adversary lower bound* for quantum query complexity, originally described in [Ambainis] and generalized many times since. Section 11 essentially generalizes the ideas in [Barnum] (a modification of the more well-known paper [Barnum-Saks-Szegedy]) for the states-of-knowledge setting; Subsection 11.2 does the same for [Yolcu]. One can see that the same symbols appearing in the semidefinite programs in [Barnum] — given different meanings — lead to the appropriate generalizations in this setup.

## 4 Sets before modding out the equivalence relation

Now to a more thorough description that includes the mixed-quantum case. With the results of the previous section, it should be easy to see that both set-ups are equivalent.

Consider an agent within an environment that could be in a state $e \in E$ (with $E$ a finite set), with some internal memory state $m \in M$. For a fixed $E$, we define sets of joint, unnormalized "generalized probability destributions", which will turn into our desired notions of "states of knowledge" after modding out equivalence relation in Section 7. We can do this for classical probability theory (probability vectors), pure quantum states (quantum wavefunctions), and mixed quantum states (quantum wavefunctions with an additional "decohering space"):[2]

$$\left.\begin{array}{c}\mathcal{S}'^{E}_{\text{class}}\\ \mathcal{S}'^{E}_{\text{quant}}\\ \mathcal{S}'^{E}_{\text{decoh}}\end{array}\right\} := \left\{\begin{array}{c}(M,P)\\ (M,P)\\ (D,M,P)\end{array}\middle| |M|,|D|<\infty, P\in\left\{\begin{array}{c}\mathbb{R}^{E\times M}_{\geq 0}\\ \mathbb{C}^{E\times M}\\ \mathbb{C}^{D\times E\times M}\end{array}\right\}\right\}.$$

$\mathbb{R}_{\geq 0}$ denotes the set of nonnegative real numbers. Note that

1. we do not fix $M$, or limit its size beyond requiring it is finite, and
2. we don't enforce normalization of our distributions.

We leave out the superscript $E$ and/or the subscript class, quant, decoh if they are irrelevant.

We define $\mathcal{S}'^{\pm} := \mathcal{S}' \times \mathcal{S}'$ (for classical, coherent-quantum, or decohering-quantum states). Denote a tuple $(S'_1, S'_2) \in \mathcal{S}'^{\pm}$ by $S'_1 - S'_2$ and interpret it as a formal difference accordingly — modulo the equivalence relation, we will call such differences "states of quasiknowledge".

As discussed in the outline, we spend the next sections defining more structure on the sets before modding out the right equivalence relation in Section 7.

## 5  Set inclusions (injective maps)

We define maps that will play the role of inclusion maps after our equivalence relation (i.e. allow us to consider some sets as subsets of other sets — will be/stay injections and form nothing but commutative diagrams when concatenated):

### 5.1  Classical and pure quantum as mixed

## quantum states

The inclusions are

1. $i_c\colon \mathcal{S}'_{\text{class}} \to \mathcal{S}'_{\text{decoh}}$,
   $i_c\left((M,P)\right) := \left(E \times M, M, \left(\sqrt{P_{e,m}}\delta_{e',e}\delta_{m',m}\right)_{(e',m'),e,m}\right)$ (in words, treating a classical as a completely decohered state),
2. $i_q\colon \mathcal{S}'_{\text{quant}} \to \mathcal{S}'_{\text{decoh}}$,
   $i_q\left((M,P)\right) := \left(\{1\}, M, (P_{e,m})_{(1,e,m),(1,e',m')}\right)$ (in words, treating a pure quantum state as a quantum state with no decoherence).

## 5.2 Knowledge as quasiknowledge

We embed $\mathcal{S}' \subset \mathcal{S}'^{\pm}$ by $i_k\colon \mathcal{S}' \to \mathcal{S}'^{\pm}$, $i_k(S') := S' - 0$.

## 5.3 Elements of E

We want to consider $e \in E$ as the state of knowledge in which the environment is guaranteed to be in state $e$, with probability 1. We map $E \to \mathcal{S}'_{\text{class}}$, $e \to (\{0\}, (\delta_{e',e})_{0,e'})$ (where $\{0\}$ is an arbitrary 1-element memory space). The situation is the same in the pure-quantum or mixed-quantum case.

## 5.4 Real numbers

$\mathbb{R}_{\geq 0} \to \mathcal{S}'_{\text{class}}$ maps $r$ to an all-$r$ matrix, again with trivial memory space. $\mathbb{R}_{\geq 0} \to \mathcal{S}'_{\text{quant}}$, $\mathcal{S}'_{\text{decoh}}$ map to an all-$\sqrt{r}$ matrix with trivial memory space.

We embed real numbers $-r' < 0$ in $\mathcal{S}'^{\pm}$ as $0 - r'$, i.e. as the negation of the embedding of $r'$.

# 6 Operations

We now define some operations on the $\mathcal{S}'$ and $\mathcal{S}'^\pm$.

## 6.1 Addition

**Addition, $+$,** is to be interpreted as the environment+agent being in either of the situations that the summands describe, and defined with **direct sums**:

1. In $\mathcal{S}'_{\text{class}}$, $(M_1, P_1) + (M_2, P_2) := (M_1 \uplus M_2, P')$, with $M_1 \uplus M_2$ denoting the disjoint union of $M_1$ and $M_2$ and the elements of $P'$ composed of the elements of $P_1$ and $P_2$.
2. In $\mathcal{S}'_{\text{quant}}$, the definition is equivalent.
3. In $\mathcal{S}'_{\text{decoh}}$,
   $(D_1, M_1, \Psi_1) + (D_2, M_2, \Psi_2) := (D_1 \uplus D_2, M_1 \uplus M_2, \Psi')$,
   with the entries of $\Psi'$ taken from $\Psi'_1, \Psi'_2$ if $d$ and $m$ belong to the same summand and set to 0 otherwise.

In the $\mathcal{S}'^\pm$, $(A - B) + (C - D) := ((A + C) - (B + D))$ as expected.

## 6.2 Multiplication

We denote **multiplication** on the $\mathcal{S}'$ by the symbol $*$, or just concatenating the factors, and interpret it as the model having access to two pieces of knowledge independently. That is:

1. In $\mathcal{S}'_{\text{class}}$, $(M_1, P_1)(M_2, P_2) := (M_1 \times M_2, P')$ with $(P')_{e,(m_1,m_2)} := (P_1)_{e,m_1}(P_2)_{e,m_2}$,
2. equivalently in $\mathcal{S}'_{\text{quant}}$,
3. and in $\mathcal{S}'_{\text{decoh}}$,
   $(D_1, M_1, \Psi_1)(D_2, M_2, \Psi_2) := (D_1 \times D_2, M_1 \times M_2, \Psi')$ with $\Psi'_{(d_1,d_2),e,(m_1,m_2)} := (\Psi_1)_{d_1,e,m_1}(\Psi_2)_{d_2,e,m_2}$.

In the $\mathcal{S}'^\pm$, $(A - B)(C - D) := (AC + BD) - (AD + BC)$.

## 6.3 Trace: Total probability

We define a linear map $\mathrm{tr}\colon \mathcal{S}'^\pm \to \mathbb{R}$ that restricts to $\mathcal{S}' \to \mathbb{R}_{\geq 0}$ as the total probability associated with a SOK: We map

1. $(M, P) \to \|P\|_{1,1}$ on $\mathcal{S}'_{\mathrm{class}}$ (where $\|P\|_{1,1}$ is the sum of entries),
2. $(M, \Psi) \to \|\Psi\|_{2,2}^2$ on $\mathcal{S}'_{\mathrm{quant}}$ (i.e. the sum of squared absolute values of entries), and
3. equivalently on $\mathcal{S}'_{\mathrm{decoh}}$.

To define the trace on $\mathcal{S}'^\pm$, we map $(S'_1 - S'_2) \to \mathrm{tr}\, S'_1 - \mathrm{tr}\, S'_2$.

## 6.4 The partial trace: Input from environment

Remember that when we consider different possible environments, we denote our sets by $\mathcal{S}'^E$ and similar. Considering a bipartite environment $E \times C$, we define a **"partial trace"** $\mathrm{tr}_C\colon \mathcal{S}'^{E \times C} \to \mathcal{S}'^E$. This works by considering the $C$ register, which previously was part of the environment, as part of the memory: it maps $M \to C \times M$, reinterprets a matrix $P \in \mathbb{R}_{\geq 0}^{(E \times C) \times M}$ as $P \in \mathbb{R}_{\geq 0}^{E \times (C \times M)}$ (and equivalently for wavefunctions) and (in the decohering case) leaves $D$ untouched. After modding out an equivalence relation in section 6, this will look more like the partial trace — after making some information accessible to the agent, the state of knowledge becomes equivalent with respect to all transformations that it can perform on that information.[3]

Of course, on $\mathcal{S}'^\pm$, $\mathrm{tr}_C(A - B) := \mathrm{tr}_C A - \mathrm{tr}_C B$.

The partial trace will be useful to discuss the interface between agent and the environment, i.e. discuss the possibility that the agent makes observations or performs actions that influence the next system state.

> After the equivalence relation of Section 7, the set inclusion $\mathbb{R} \subseteq \mathcal{S}^{\pm\{0\}}$ defined in Subsection 5.4 will be a *surjective* vector space isomorphism for a 1-element $E = \{0\}$. Under that isomorphism, the result of treating an environment $E$ as $\{0\} \times E$ and calculates the partial trace over $E$ will be consistent with the trace of Subsection 6.3, as we expect from quantum physics.

## 6.5 The preorder

We define a **preorder** $S'_1 \leq S'_2$ (i.e. a relation such that $S'_1 \leq S'_1$, and $S'_1 \leq S'_2$, $S'_2 \leq S'_3$ implies $S'_1 \leq S'_3$) to capture the notion that the agent can trivially transform a state into another state as follows:

1. In $\mathcal{S}'_{\text{class}}$, $(M_1, P_2 T^T) \leq (M_2, P_2)$ for all transformation matrices on the memory $T \in \mathbb{R}_{\geq 0}^{M_1 \times M_2}$ with $\|T\|_1 \leq 1$, where
   - $P_2 T^T$ is the usual matrix product of $P_2$ with the transpose of $T$: One can check that this corresponds to applying $T$ on the $M$ subsystem; this is a technicality related to not using graphical notation.
   - The norm is the 1-norm of $T$, i.e. the maximal column sum. This means that starting from any state $m_2 \in M_2$, the probabilities of transitioning to a final state must sum to **at most** one; if it is $< 1$ for some column, we interpret this as the agent "giving up" (and outputting "failure") with a certain probability. If all column sums are 1, we have a "proper" transition matrix that doesn't lose probability mass.

2. The definition in $\mathcal{S}'_{\text{quant}}$ is analogous (with $\mathbb{R}_{\geq 0}$ replaced by $\mathbb{C}$, but the transpose not replaced by a Hermitian conjugate) except that our constraint on $T$ is that $\|T\|_2 \leq 1$, using the 2-norm of $T$. This means that $T$ is a contraction w.r.t. the 2-norm of states it is applied on, or equivalently, every singular value is at most 1. Analogously to 1., *all* singular values being 1 is equivalent to $T$ being an isometry, i.e. a "proper" quantum transition matrix. If some singular values of $T$ are $< 1$, the block-matrix

$$\begin{pmatrix} T \\ \sqrt{I - T^\dagger T} \end{pmatrix}$$

is an isometry nevertheless; we can imagine the agent gives up if it lands in the lower block of that matrix and/or consider $T$ to be one [Kraus operator of a quantum channel](#) [NielsenChuang].

3. In $\mathcal{S}'_{\text{decoh}}$, we consider transitions involving both $D$ and the memory. So the condition is

$$(D_1, M_1, T_D(\Psi_2 T_M^T)) \leq (D_2, M_2, \Psi_2),$$

where $T_D(\Psi_2 T_M^T)$ denotes

1. The application of some $T_M \in \mathbb{C}^{(M_1 \times D_M) \times M_2}$ to the memory space of $\Psi_2$, where $D_M$ is an intermediate decohering space and $\|T_M\|_2 \leq 1$,
2. following that, the application of $T_D \in \mathbb{C}^{D_1 \times (D \times D_M)}$ to $D$ and $D_M$, with $\|T_D\|_2 \leq 1$ as well. So $T_M$ transforms the memory and generates a register $D_M$ that becomes part of the decohering space, and $T_D$ is an arbitrary transformation of that decohering space, which is **not necessarily an isometry either**.[4]

On $\mathcal{S}'^\pm$, $(A - B) \leq (C - D)$ iff $A + D \leq B + D$. On $\mathcal{S}'$, we can easily see that the preorder axioms are fulfilled. For $\mathcal{S}'^\pm$, we'll show this in Subsection [7.1](#).

## 6.6 Sub-preimage of the partial trace: Output to environment

The partial trace corresponds to the agent incorporating part of the environment's state into its internal memory. Conversely, suppose the agent is in state $S_0' \in \mathcal{S}'^E$. Then the states $S_X' \in \mathcal{S}'^{E \times X}$ that it can obtain by applying an acceptable internal memory transformation, and then outputting a part of its internal memory into an output register $X$, are given by the set

$$\left\{ S'_X \mid \mathrm{tr}_X S'_X \leq S'_0, \; S'_X \in \mathcal{S}'^{E \times X} \right\}.$$

# 7 Modding out

This section contains the technical details of this article's central claim that $\mathcal{S} \subseteq \mathcal{S}^{\pm}$, defined by modding out the equivalence relation given by the condition $S'_1 \leq S'_2 \leq S'_1$, is a **convex cone as a subset of an associative, commutative $\mathbb{R}$-algebra that is ordered w.r.t. $\leq$ as a vector space, and the partial trace $\mathrm{tr}_C$ is a monotonous vector space homomorphism that is surjective when restricted to $\mathcal{S}^{E \times C} \to \mathcal{S}^E$.**

## 7.1 The cancellation property

The following property is important. The proof of the harder direction is similar to the universal query algorithm in Subsection 11.3.

> **Theorem: Cancellation property**
>
> Let $S'_1, S'_2, S'_3 \in \mathcal{S}'^{\pm}$. Then $S'_1 + S'_3 \leq S'_2 + S'_3$ if and only if $S'_1 \leq S'_2$.

> **Proof**
>
> We prove the statement for $S'_1, S'_2, S'_3 \in \mathcal{S}'$, generalizing to $\mathcal{S}'^{\pm}$ is easy after we have shown that $\mathcal{S}'^{\pm}$ is in fact a preorder.
>
> If $S'_1 \leq S'_2$ by some transformation $P$, we can construct a transformation witnessing $S'_1 + S'_3 \leq S'_2 + S'_3$ by applying $P$ to $S'_2$ and identity to $S'_3$.

> The converse is more interesting. First, note that for any $A, B, C \in \mathcal{S}'$,

$$A + B \leq B + A$$

and

$$(A + B) + C \leq A + (B + C)$$

by operations that relabel the indices. Furthermore, note that $NS' \leq \sum_{i=1}^{N} S' \leq NS'$ for all $S' \in \mathcal{S}'$ and integers $N \geq 1$.

Now suppose that $S_1' + S_3' \leq S_2' + S_3'$. By the last two paragraphs, this implies that $S_1' + S_3' \leq S_3' + S_2'$, and we can ignore brackets. Applying this inequality $N$ times, we obtain

$$NS_1' + S_3' \leq \sum_{i=0}^{N-1} S_1' + S_3' \leq S_3' + \sum_{i=0}^{N-1} S_2' \leq S_3' + NS_2' \leq NS_2' + S_3'.$$

After rescaling, we obtain $S_1' + \frac{S_3'}{N} \leq S_2' + \frac{S_3'}{N}$ for any positive integer $N$. Applying the transformation witnessing this inequality to $S_2'$, we obtain a state arbitrarily close to $S_1'$ with respect to a reasonable metric. Therefore, $S_1'$ is reachable as well by a completeness argument:

1. For given memory spaces, the set of states $S'$ achievable from $S_2'$ is the image of the (continuous) map that maps the (compact) set of transformations to the result of evaluating them at $S_2'$,
2. therefore it is compact,
3. a compact metric space is complete,
4. so the limit of the Cauchy sequence approaching $S_1'$ is achievable as well.

This allows us to show that $\leq$ on the $\mathbb{S}'^{\pm}$ as defined in Subsection 6.5 is in fact a preorder:

1. Reflexivity is trivial,

2. for transitivity, consider $(A - B) \leq (C - D)$ and
   $(C - D) \leq (E - F)$. Then, using preorder transitivity in $\mathcal{S}'$,
   1. $A + D \leq C + B$, therefore $A + E + D \leq B + E + C$,
   2. $C + F \leq E + D$, therefore
      $A + F + C \leq A + E + D \leq B + E + C$,
   3. By the cancellation property, $A + F \leq B + E$, i.e.
      $(A - B) \leq (E - F)$.

## 7.2 The equivalence relation and algebra

Finally, we define an **equivalence relation** on the $\mathcal{S}'$ and $\mathcal{S}'^{\pm}$ by $X \sim Y \leftrightarrow (X \leq Y \leq X)$, and call the equivalence classes the spaces of **states of knowledge** $\mathcal{S}$ and **states of quasiknowledge** $\mathcal{S}^{\pm}$. As promised, a reader familiar with Grothendieck group constructions may recognize the latter as an instance. We can easily show this is an equivalence relation on the $\mathcal{S}'$, and using the cancellation property of the last section, we can show that it is an equivalence relation on $\mathcal{S}'^{\pm}$ as well.

It is tedious but mostly straightforward to check that all our operations behave well with this equivalence relation and the claim in the beginning of Section 7, that $\mathcal{S} \subseteq \mathcal{S}^{\pm}$ becomes a convex cone as a subset of an associative, commutative $\mathbb{R}$-algebra that is ordered w.r.t. $\leq$ as a vector space. This means:

1. For all operations, choosing different representatives of the same equivalence class leads to equivalent results,
2. $\mathcal{S}^{\pm}$ is a commutative ring, fulfilling the ring axioms enumerated, for example, in Wikipedia, as well as the axiom that $*$ is commutative. Copying the content of the link, for any $A, B, C \in \mathcal{S}^{\pm}$,
   1. $(\mathcal{S}^{\pm}, +)$ is an Abelian (commutative) group, i.e.
      1. $+$ is associative, $A + (B + C) = (A + B) + C$,
      2. $+$ is commutative (abelian), $A + B = B + A$,
      3. There is a neutral element 0 such that $A + 0 = A$,

4. each $A$ has an inverse $-A$ such that $A + (-A) = 0$.
   2. $(\mathcal{S}^{\pm}, *)$ is a monoid with neutral element 1, i.e.
        1. $*$ is associative as above,
        2. $*$ is commutative as above (this is what is meant by "commutative ring"),
        3. $1 * a = a * 1 = a$ for all $a \in \mathcal{S}^{\pm}$.
   3. the distributive law holds, i.e. $A(B + C) = (AB) + (AC)$; $(A + B)C = (AC) + (BC)$.
3. $\mathcal{S}^{\pm}$ is an associative $\mathbb{R}$-algebra, meaning that the embedding map $\mathbb{R} \to \mathcal{S}^{\pm}$ is a ring homomorphism ($f(a + b) = f(a) + f(b)$, $f(1) = 1$ and $f(ab) = f(a)f(b)$); furthermore, the image lies in the center (i.e. for any $r \in \mathbb{R}$, $f(r)$ commutes with all $S \in \mathcal{S}^{\pm}$). These follow from the definitions.
4. The map $\mathcal{S} \to \mathcal{S}^{\pm}$, $s \to (s, 0)$ is injective (which is necessary to be able to consider $\mathcal{S}$ as a subset of $\mathcal{S}^{\pm}$). This follows from the cancellation property of Subsection 7.1, as
$A - 0 \leq B - 0 \leq A - 0$ implies that $A + 0 \leq B + 0$, and $B + 0 \leq A + 0$. By the cancellation property, this in turn implies $A \leq B \leq A$, i.e. $A = B$.
5. $\mathcal{S}^{\pm}$ is ordered w.r.t. $\leq$ as a vector space. This means that $S_1 \leq S_2$ implies $S_1 + S_3 \leq S_2 + S_3$, and $S_1 \leq S_2$ implies $\lambda S_1 \leq \lambda S_2$ for any $\lambda \in \mathbb{R}_{\geq 0}$. This can be verified easily, again using the cancellation property.

> **Question/Remark**
>
> Is it an ordered algebra as well? That seems to be true for $\mathcal{S}^{\pm}_{\text{quant}}$ and $\mathcal{S}^{\pm}_{\text{class}}$ (see Section 3), but I don't know whether the product of positive states is always positive in the faulty-quantum situation.
>
> Using the equivalence discussed here, it may be more helpful to define the $S_1 \leq S_2$ by $S_2 - S_1 \in \mathcal{S}$.

6. $\mathcal{S}$ is a convex cone: $\mathcal{S}$ is closed under convex combinations and scalar multiplications.

The partial trace $\text{tr}_C\colon \mathcal{S}^{\pm(E\times C)} \to \mathcal{S}^{\pm E}$ is a **linear map**[5] that surjectively maps states of knowledge to states of knowledge (i.e. the image $\text{tr}_C(\mathcal{S}^{E\times C}) = \mathcal{S}^E$). Conversely, however, the preimage of a SOK *does* generally contain non-SOKs. $\text{tr}_C$ is also monotonous in $\leq$.

## 8 Two examples

These two examples are supposed to illustrate this approach. The second one, of Subsection 8.2, introduces a new concept as well: Formal power series and differential equations of knowledge.

### 8.1 Determining the bias of a coin

God has chosen a biased (classical) coin which either shows heads with $p = 0.6$ or tails with $p = 0.6$, the prior probabilities for each are equal. The "agent" (in this simple example) can't make any decisions to influence the environment, it only observes a sequence of coin flips.

The environment is a 2-element set $E = \{\text{HeadsBiased}, \text{TailsBiased}\}$; denote this by $E = \{0, 1\}$ for simplicity. One representative for the initial state of knowledge $S_0$ is $(M, P)$ with $M = \{0\}$ and $P = \begin{pmatrix} 1/2 \\ 1/2 \end{pmatrix} = \frac{1}{2}\mathbf{1}$. This corresponds to the agent having a trivial internal memory.

The agent could perform some internal memory transformations. For example, it could introduce a 3-state memory $M' = \{0, 1, 2\}$, and randomly choose $0$ with probability $1/2$, and $1$ or $2$ with probability $1/4$ each. Then we'd get a representative $(M', P')$ with $P' := \begin{pmatrix} \frac{1}{4} & \frac{1}{8} & \frac{1}{8} \\ \frac{1}{4} & \frac{1}{8} & \frac{1}{8} \end{pmatrix}$. By our definitions, these representatives are equivalent as states of knowledge:

$$P' = PT^T, \ T := \begin{pmatrix} \frac{1}{2} \\ \frac{1}{4} \\ \frac{1}{4} \end{pmatrix},$$

$$P = P'T'^T, \ T' := \begin{pmatrix} 1 & 1 & 1 \end{pmatrix},$$

and both $T$ and $T'$ have 1-norm (maximal column sums) $\leq 1$. The latter transformation matrix corresponds to forgetting $M'$ and getting back to a trivial memory state. So as we hoped, we have modded out equivalent transformations that the agent could perform on its internal memory.

The situation becomes more interesting after the agent made a few observations of coin tosses. Suppose it makes a single observation and stores the result as a memory state in $M := \{h, t\}$. Then the joint probability matrix becomes

$$P := \begin{pmatrix} \frac{0.6}{2} & \frac{0.4}{2} \\ \frac{0.4}{2} & \frac{0.6}{2} \end{pmatrix},$$

and a representative of our state of knowledge is $(M, P)$.

After it made 2 observations (and stored both results), a representative is $(M_2, P_2)$ with $M_2 := \{hh, ht, th, tt\}$ and

$$P_2 := \begin{pmatrix} \frac{0.6*0.6}{2} & \frac{0.6*0.4}{2} & \frac{0.4*0.6}{2} & \frac{0.4*0.4}{2} \\ \frac{0.4*0.4}{2} & \frac{0.4*0.6}{2} & \frac{0.6*0.4}{2} & \frac{0.6*0.6}{2} \end{pmatrix}:$$

The internal memory can store all combinations of outcomes, and standard probability theory yields the joint probabilities.

We know that it is superfluous to store all outcomes: It suffices to keep track of *how many times* the coin showed heads, as the order doesn't give additional information about which coin was chosen. We will now see that the formalism reflects this, as a representative for the latter form of storage is equivalent as a state of knowledge: When counting the coin flips, the joint probability matrix for observing a given numbers of heads is given by binomial distributions, i.e. we get a representative $(\{0, 1, 2\}, P_2')$ with

$$P_2' := \begin{pmatrix} 0.18 & 0.24 & 0.08 \\ 0.08 & 0.24 & 0.18 \end{pmatrix}.$$

We find transformation matrices with column sums $\leq 1$ such that $P_2' = P_2 T_2^T$ and $P_2 = P_2' T_2'^T$, proving that these representatives are equivalent:

$$T_2 := \begin{pmatrix} 1 & 0 & 0 & 0 \\ 0 & 1 & 1 & 0 \\ 0 & 0 & 0 & 1 \end{pmatrix},$$

$$T_2' := \begin{pmatrix} 1 & 0 & 0 \\ 0 & 0.5 & 0 \\ 0 & 0.5 & 0 \\ 0 & 0 & 1 \end{pmatrix}.$$

$T_2$ corresponds to counting the number of heads; $T_2'$ corresponds to randomly generating an order of flips given a count.

In general, given a SOK $S \in \mathcal{S}$, observing an additional coin flip and storing the result corresponds to multiplying $S$ by a SOK $Q$ represented by $(M_Q, P_Q) := \left( \{h, t\}, \begin{pmatrix} 0.6 & 0.4 \\ 0.4 & 0.6 \end{pmatrix} \right)$ according to our formalism. In other words, if $S$ is represented by $(M, P)$, it gets mapped to the equivalence class of $(M \times M_Q, P')$ with $P'_{e,(m,x)} := P_{e,m}(P_Q)_{e,x}$. So after $n$ coin flips, the state of knowledge is

$$S = Q^n S_0.$$

Finally, suppose that in each step, the agent only observes a coin flip with probability 1/4, and doesn't change its state of knowledge otherwise. So it gets one of the outcomes $M_p := \{h, t, n\}$, with a conditional probability matrix

$$\begin{pmatrix} 0.15 & 0.10 & 0.75 \\ 0.10 & 0.15 & 0.75 \end{pmatrix} = \frac{1}{4}\begin{pmatrix} 0.6 & 0.4 \\ 0.4 & 0.6 \end{pmatrix} \oplus \frac{3}{4}\begin{pmatrix} 1 \\ 1 \end{pmatrix} = \frac{1}{4}Q + \frac{3}{4}\mathbf{1}.$$

Here, the $\oplus$ signs stand for concatenating the columns, just like in our definition of addition. Then the state of knowledge after $n$ events (that are coin flips with probability 1/4) is

$$\left(\frac{1}{4}Q + \frac{3}{4}\mathbf{1}\right)^n S_0.$$

Similarly as in the situation above, one can show that it's sufficient for the agent to store a counter of heads and tails observed so far.

## 8.2 Differential equation of knowledge

Together, addition and multiplication allows us to define formal power series of knowledge (we leave mathematical rigour for a moment, as we didn't define infinite sums or limits). For example, suppose our learner observes the stars with a telescope without making any choices. In an infinitesimal time $\Delta t \to 0$, it observes a supernova with probability $r\Delta t$, generating experimental data $A \in \mathcal{S}$. Otherwise, it observes nothing. If $S(t)$ is the state of knowledge over time, we obtain

$$S(t + \Delta t) \to ((1 - r\Delta t)\mathbf{1} + r\Delta t A)S(t).$$

This is solved by

$$S(t) = \exp((A - \mathbf{1})rt)S(0),$$

which is to be interpreted as the appropriate formal infinite power series. In fact, writing

$$K(t) = \exp(-rt)\exp(Art)S(0) = \sum_{k=0}^{\infty} \frac{\exp(-rt)(rt)^k}{k!} A^k S(0)$$

shows that the amount of knowledge obtained follows a Poisson distribution. Note that these equations contain minus signs, so our Grothendieck construction was probably helpful.

## 9 Convex optimization

We have defined the algebraic structure and given two concrete examples. The next two sections will be devoted to understanding the "statics" and "dynamics" of states of knowledge better. The partial trace will play a central role in these.

All of these conditions will be **convex** conditions, and convex optimization problems, because the SOKs are convex and partial traces are linear. So when we **truncate to a finite basis, we are able to use convex optimization and duality** to find optimal algorithms that allow the agent to determine some value or perform some manipulation of the environment. The SDPs obtained in that way are mostly generalizations of [Barnum-Saks-Szegedy], [Barnum] (which discussed this question for quantum wavefunctions), and the adversary bound - quantum query algorithm duality as presented in [Yolcu].

## 10 Output conditions

We now define some "output conditions" on SOKs, answering the questions

- "If an algorithm achieves a certain SOK, then how close is it to a target SOK?", and
- "Given a SOK, what functions can the algorithm calculate with a given error probability?"

## 10.1 Measurements

Given $S \in \mathcal{S}^E$, we define $\text{eval}(S) \in \mathbb{R}_{\geq 0}^E$ by

$$(\text{eval}(S))_e := \text{tr}(eS).$$

In words, eval maps $S$ to a nonnormalized probability distribution over environmental states.

Now suppose that the agent should output some information $c \in C$. As in Subsection 6.6, the set of joint (nonnormalized) probability distributions over $E$ and $C$ that the agent is able to output is

$$\left\{ \text{eval}(S_C) \mid \text{tr}_C S_C \leq S, \ S_C \in \mathcal{S}^{E \times C} \right\} \subset \mathbb{R}_{\geq 0}^{E \times C}.$$

This is clearly a convex set, as promised.

It is easy to turn this definition into optimization problems to find optimal strategies to guess properties of the environment, or sets of states of knowledge that allow such guesses with certain success rates. For example, suppose that $S$ is normalized, and define a "utility function" $V \in \mathbb{R}^{E \times C}$ — a matrix of payoffs if the environmental state is $E$, and the algorithm outputs $C$. For example, for some function $f: E \to C$, $V[e, c] = \delta_{e, f(c)}$ would correspond to the agent succeeding if it correctly determines this function of the environment. Then solving the optimization problem

$$\begin{aligned}
\underset{S_C \in \mathcal{S}^{C \times E}}{\text{maximize}} \quad & \text{eval}(S_C) \cdot V \\
\text{subject to} \quad & \text{tr}_C S_C \leq S,
\end{aligned}$$

with $\cdot$ the elementwise inner product between these matrices, corresponds to maximizing the expected payoff. On the set of states of knowledge, the constraint that this optimization problem can be solved with value at least $\lambda^*$ is

$$\left\{ S \mid \exists S_C \in \mathcal{S}^{C \times E} : \lambda^* \leq \mathrm{eval}(S_C) \cdot V, \mathrm{tr}_C S_C \leq S, S \in \mathcal{S}^E \right\}.$$

This is a convex set.

To maximize the payoff of the worst-case inputs, we instead consider the optimization problem

$$\begin{aligned}
\underset{S_C \in \mathcal{S}^{C \times E}}{\text{maximize}} \quad & \min_{e \in E} \mathrm{eval}(S_C) \cdot (e^T e V) \\
\text{subject to} \quad & \mathrm{tr}_C S_C \leq S,
\end{aligned}$$

where $e^T e V$ denotes $V$ with all rows except the one corresponding to $e$ replaced by zeroes.

$$\begin{aligned}
\underset{S_C \in \mathcal{S}^{C \times E}, \lambda \in \mathbb{R}_{\geq 0}}{\text{maximize}} \quad & \lambda \\
\text{subject to} \quad & \lambda \leq \mathrm{eval}(S_C) \cdot (e^T e V) \qquad \forall e \in E, \\
& \mathrm{tr}_C S_C \leq S,
\end{aligned}$$

which is evidently a convex optimization problem that is equivalent to a semidefinite program in the pure quantum case. We can turn this optimization problem into a convex constraint on the set of states of knowledge in analogy to the average-case situation above.

## 10.2 Quantifying similarity: Trace distance

We generalize the SDP that quantifies the trace distance of quantum states by defining, for $S, T \in \mathcal{S}^E$, $\|S - T\|_{\mathrm{tr}}$ as the infimal value of the optimization problem

$$\begin{aligned}
\underset{\Delta \in \mathcal{S}^E}{\text{minimize}} \quad & \mathrm{tr}\,\Delta \\
\text{subject to} \quad & -\Delta \leq S - T \leq \Delta.
\end{aligned}$$

One can show easily that this is a metric.

# 11 An agent interacting with its environment

## 11.1 Evolution

Now consider an agent interacting with its environment. So we introduce some output and input subsystems $O$, $I$, and a "law of nature" $T\colon \mathcal{S}^{E\times O} \to \mathcal{S}^{E\times I}$ that evolves the environment influenced by the agent's choices, and generates measurement data. The "law of nature" should be instantiated as some transition matrix $T'$, similar to the ones in our order definition in Subsection 6.5:

1. For $\mathcal{S}_{\text{class}}$, $T' \in \mathbb{R}_{\geq 0}^{(E\times I)\times(E\times O)}$ and $\|T'\|_1 \leq 1$,
2. For $\mathcal{S}_{\text{quant}}$, $T' \in \mathbb{C}^{(E\times I)\times(E\times O)}$ and $\|T'\|_2 \leq 1$,
3. For $\mathcal{S}_{\text{decoh}}$, $T' \in \mathbb{C}^{(D_T\times E\times I)\times(E\times O)}$ and $\|T'\|_2 \leq 1$, i.e. $T'$ may add some new space $D_T$ to the decohering space.

We can clearly form a tensor products of transformations from the product of their constituting matrices; denote this by $T_1 \otimes T_2$ as expected.

Then, starting with a SOK $S \in \mathcal{S}^E$, the states of knowledge attainable by the agent in one step are

$$\left\{\text{tr}_I(T(S_O)) \mid \text{tr}_O S_O \leq S, S_O \in \mathcal{S}^{E\times O}\right\}:$$

As in Subsection 6.6, if a state $S_O \in \mathcal{S}^{E\times O}$ can be obtained by an agent writing into the output register, based on $S \in \mathcal{S}^E$, then reincorporating $O$ in $E$ (i.e. applying the partial trace) would need to yield a state $\leq \mathcal{S}^E$. Conversely, all states fulfilling this can be obtained. Then the transformation corresponds to applying $T$, and incorporating the next step's input into the internal memory is the same as application of the partial trace.

## 11.2 The adversary bound

Consider a multi-step evolution with each step as in Subsection 11.1: Starting from $0 \neq S_0 \in \mathcal{S}^E$, the system goes through intermediate states $S_1, S_2, \ldots, S_{N-1}$ and arrives at a final state $S_N$ — influenced by the agent, whose choices in the different steps are encoded by SOKs $S_{O,k} \in \mathcal{S}^{E \times O}$:

$$S_{k+1} = \mathrm{tr}_I(T(S_{O,k})),$$
$$\mathrm{tr}_O S_{O,k} \leq S_k, S_{O,k} \in \mathcal{S}^{E \times O}$$

with $k \in \{0, 1, \ldots N-1\}$. By linearity, we can add these expressions and obtain

$$\sum_{k=1}^{N} S_k = \mathrm{tr}_I\left(T\left(\sum_{k=0}^{N-1} S_{O,k}\right)\right),$$
$$\mathrm{tr}_O\left(\sum_{k=0}^{N-1} S_{O,k}\right) \leq \sum_{k=0}^{N-1} S_k.$$

We add $S_0$ on both sides of the equality, and $S_N$ on both sides of the inequality. Then we can combine these expressions:

$$S_N + \mathrm{tr}_O\left(\sum_{k=0}^{N-1} S_{O,k}\right) \leq \sum_{k=0}^{N} S_k = S_0 + \mathrm{tr}_I\left(T\left(\sum_{k=0}^{N-1} S_{O,k}\right)\right).$$

Now let $\widetilde{S} := \sum_{k=0}^{N-1} S_{O,k}$. Then $\widetilde{S} \in \mathcal{S}$. Furthermore, by induction, all $\mathrm{tr} S_{O,k} \leq \mathrm{tr} S_0$ — implying $\mathrm{tr} \widetilde{S} \leq N \mathrm{tr} S_0$, or $N \geq \mathrm{tr}\widetilde{S}/\mathrm{tr}S_0$ (if $S_0 \neq 0$, $\mathrm{tr}S_0 > 0$ and the division is permissible).

> ⓘ A bit of intuition: Up to multiplying by $1/N$, $\widetilde{S}$ corresponds to applying a "random step" of the algorithm.

In particular, any algorithm performing the transformation $S_0 \to S_N$ must take at least

$$N \geq \left\lceil \frac{\mathrm{Adv}(S_N - S_0)}{\mathrm{tr} S_0} \right\rceil$$

steps, where we define $\mathrm{Adv}(S_N - S_0)$ as the infimal value of the optimization problem

$$\begin{aligned}
\underset{\widetilde{S} \in \mathcal{S}^{E \times O}}{\text{minimize}} \quad & \mathrm{tr} \widetilde{S} \\
\text{subject to} \quad & \mathrm{tr}_I \left( T \left( \widetilde{S} \right) \right) - \mathrm{tr}_O \widetilde{S} \geq S_N - S_0.
\end{aligned}$$

Note that Adv only depends on $S_N - S_0$, and the scale invariance: For any $\lambda \in \mathbb{R}_{\geq 0}$,

$$\mathrm{Adv}(\lambda(S_N - S_0)) = \lambda \mathrm{Adv}(S_N - S_0).$$

We strengthen this bound in case the transformation matrix that gives rise to $T$ is block-diagonal in $E$ (i.e. applying it never changes the environmental state — this means that the agent is a "query computer" that can only observe the environment, but not change it). In this case, it is not only true that

$$\mathrm{tr} \widetilde{S} \leq N \mathrm{tr} S_0,$$

but that

$$\mathrm{tr} \left( e \mathrm{tr}_O \widetilde{S} \right) \leq N \mathrm{tr} \left( e S_0 \right)$$

for each $e \in E$. This can be seen by induction. So under the assumption that $\mathrm{tr} \left( e S_0 \right) = 1$ for all $e \in E$, an alternative, stronger lower bound is

$$\begin{aligned}
\underset{\widetilde{S} \in \mathcal{S}^{E \times O}}{\text{minimize}} \quad & \max_{e \in E} \mathrm{tr} \left( e \mathrm{tr}_O \widetilde{S} \right) / \mathrm{tr} \left( e S_0 \right) \\
\text{subject to} \quad & \mathrm{tr}_I \left( T \left( \widetilde{S} \right) \right) - \mathrm{tr}_O \widetilde{S} \geq S_N - S_0.
\end{aligned}$$

We call this bound $\widetilde{\mathrm{Adv}}(S_N - S_0)$.

> **Side remark**
>
> We could also slightly strengthen this bound by adding the constraint
>
> $$\mathrm{tr}_O\left(\widetilde{S}\right) - S_0 \in \mathcal{S},$$
>
> which must be true for $\widetilde{S}$ obtained from algorithms as well. I didn't put this constraint in by default, as it would be irrelevant for $\mathrm{Adv}(S_N - S_0) \gg 1$, make $\mathrm{Adv}$ dependent on $S_0$ explicitly (rather than just $S_N - S_0$) and break the scale invariance.

## 11.3  The universal query algorithm

Suppose that the algorithm can always choose to do nothing, i.e.

$$\forall S \in \mathcal{S}^E \, \exists S_{O,\mathrm{idle}} \in \mathcal{S}^{E \times O} : \mathrm{tr}_O S_{O,\mathrm{idle}} = \mathrm{tr}_O\left(T\left(S_{O,\mathrm{idle}}\right)\right) = S.$$

Under this condition, we can turn the feasible solutions of the optimization problems specifying the adversary bound above into algorithms approximating the desired state transformation. This works similarly as the proof of the cancellation property in Subsection 7.1, as mentioned there.

Consider a feasible solution for the optimization problem corresponding to $\mathrm{Adv}(S_R - S_S)$, let the idle states corresponding to initial and target states be above states be $S_{O,S}$ and $S_{O,R}$. Choose a number of steps $N'$. Then the algorithm given by the intermediate states of knowledge

$$S_{O,k} := \frac{N' - k}{N'} S_{O,S} + \frac{k}{N'} S_{O,N} + \frac{\widetilde{S}}{N'}, \; 0 \le k < N'$$

results in a valid query algorithm as in the last Subsection 11.2, transforming

$$S_S + \frac{\widetilde{S}}{N'} \to S_R + \frac{\widetilde{S}}{N'}$$

in $N'$ steps. Finally, we can transform the result to $S_N$ by a non-invertible transformation; as we let $N' \to \infty$, this approaches an algorithm that transforms $S_S \to S_R$ without error.

> (i) To quantify how close to the target SOK the algorithm allows us to get starting from $S_S$, we can unroll the definitions and work with the representatives of the states of knowledge. Then the algorithms correspond to sequences of linear transformation that don't increase the $L_1$ norm (for classical SOK)/squared-$L_2$ norm (for pure/mixed-quantum SOK). There exist representative of $S_S$ and $S_S + \frac{\widetilde{S}}{N'}$ whose $L_1$/squared-$L_2$ distance from each other is at most $\operatorname{tr}\widetilde{S}/N'$. Applying the algorithm intended for the representative of $S_S + \frac{\widetilde{S}}{N'}$ to the representative of $S_S$ instead will yield a final state whose $L_1$/squared-$L_2$-distance from $S_R$ is at most $\operatorname{tr}\widetilde{S}/N'$.
>
> If we do the same with a witness to $\widetilde{\mathbf{Adv}}$ (found for a problem which meets the block-diagonality assumption), we can apply this argument to each possible input state individually, i.e. for any $e \in E$, applying the algorithm to $eS_0$ will lead to an output state $eS'_N$ which has the same distance bound as above. This is the usual error bound for universal query algorithms. See [Yolcu] for a more systematic description.
>
> I am not satisfied with this description and hope there is a more elegant error quantification that doesn't require going back to representatives. In the quantum case, we know by Uhlmann's theorem [NielsenChuang] that the fidelity is related to the minimal achievable $L_2$-distance between representations of states of knowledge. I hope one can generalize this and relate it to the convex optimization

> formulation somehow. See [Yolcu] for details on the proof for pure quantum states.

> If $\widetilde{S}$ was constructed as a random step of an algorithm as above, we can interpret the algorithms constructed from $\widetilde{S}$ as "factory line productions" of $S_N$ from $S_0$. Adding $\widetilde{S}$ is equivalent to adding all intermediate states of the original algorithm at once, and one step of the constructed algorithm corresponds to applying one processing step to each of the intermediate states.

## 12 Discussion and outlook

I see the following main open questions and directions:

1. Can we apply this theory to something less trivial — for example, lower bounds for concrete faulty quantum query algorithms? I see a possibility that one can use it to prove a theorem of the sort "if one can solve a problem using a faulty quantum computer in time T, one can also solve it using a classical/pure-quantum computer in time T'" — by converting optimal adversary states of knowledge into each other or similar.

2. How to understand the convex optimization duals better, prove an analogue of Slater's condition, and parametrize the dual of the vector space of SOKs similarly to how one can parametrize the dual of the space of Hermitian matrices by Hermitian matrices themselves? Relatedly, defining an inner product on the vector space of states of knowledge is a natural question. In the pure-quantum case described as in Subsection 3.2, the standard inner product of Hermitian matrices corresponds to mapping $[\Psi_1] \cdot [\Psi_2] \to |\langle \Psi_1 | \Psi_2 \rangle|^2$, and extending bilinearly. A candidate for a classical analogue would be the squared Bhattacharyya

coefficient $\left[\vec{p_1}\right] \cdot \left[\vec{p_2}\right] \to \left(\sum_{e \in E} \sqrt{(p_1)_e (p_2)_e}\right)^2$, but the resulting properties are not clear to me.

3. Can one prove strong direct product theorems in this formalism in the sense of Lee-Roland? The "Lee-Roland fidelity" as defined in [Lee-Roland], equation 3.2, was a central technical tool in that paper, and in the pure-quantum case, it has a semidefinite programming formulation.

4. In analogy of the formal power series of Subsection 8.2, one may be able to describe formal power series if the agent is allowed to make choices — when considering *sets of attainable states of knowledge* instead of *individual states of knowledge*.

5. This document does not contain my attempt to generalize an SDP defining the fidelity of quantum states. This is because the "right" way to generalize the quantity $\text{tr}(X\rho)$, with $\rho$ a pure-quantum SOK over a qubit, interpreted as a density matrix, is not clear to me. Similarly, one can define the similarity measure used in the error quantification of Subsection 11.3 in more detail. As mentioned, in the pure-quantum case, Uhlmann's theorem [NielsenChuang] relates this similarity measure to the fidelity, and it is desirable to generalize fidelity in such a way that Uhlmann's theorem generalizes as well.

## 14 Footnotes

[1]: Our convention is that each row is a different environmental state, and each column a different internal memory state.

[2]: I think that using density matrices and a variant of quantum channels would work as well to discuss mixed states.

[3]: In the correspondence between $\mathcal{S}_{\text{quant}}^{E}$ and reduced density matrices on $E$, this works out to be a literal partial trace.

[$^4$]: I think that one can obtain a more elegant description of possible transformations by adapting the parallel developments by [Gutoski-Watrous] and [Chiribella-D'Ariano-Perinotti]. But I don't see a use of that as of this note.

[$^5$]: It is **not** a homomorphism of $\mathbb{R}$-algebras.

*formatted by Markdeep 1.14*